\documentclass[twocolumn]{aastex62}

\defcitealias{Barone2018}{Paper I}

\graphicspath{{./}{figures/}}

\received{\today}
\revised{---}
\accepted{---}

\submitjournal{ApJ}

\shorttitle{Star-Forming Stellar Populations}
\shortauthors{Barone et al.}

\begin{document}

\title{Gravitational Potential and Surface Density Drive Stellar Populations --- II. Star-Forming Galaxies}

\correspondingauthor{Tania M.\ Barone}
\email{tania.barone@anu.edu.au}

\author[0000-0002-2784-564X]{Tania M.\ Barone}
\affiliation{Research School of Astronomy and Astrophysics, The Australian National University, Canberra, ACT 2611, Australia}
\affiliation{Sydney Institute for Astronomy, School of Physics, The University of Sydney, NSW, 2006, Australia}
\affiliation{ARC Centre of Excellence for All Sky Astrophysics in 3 Dimensions (ASTRO 3D)}
\author[0000-0003-2388-8172]{Francesco D'Eugenio}
\affiliation{Sterrenkundig Observatorium, Universiteit Gent, Krijgslaan 281 S9, B-9000 Gent, Belgium}
\author[0000-0001-9552-8075]{Matthew Colless}
\affiliation{Research School of Astronomy and Astrophysics, The Australian National University, Canberra, ACT 2611, Australia}
\affiliation{ARC Centre of Excellence for All Sky Astrophysics in 3 Dimensions (ASTRO 3D)}
\author[0000-0001-8495-8547]{Nicholas Scott}
\affiliation{Sydney Institute for Astronomy, School of Physics, The University of Sydney, NSW, 2006, Australia}
\affiliation{ARC Centre of Excellence for All Sky Astrophysics in 3 Dimensions (ASTRO 3D)}

%~~~~~~~~~~~~~~~~~~~~~~~~~~~~~~~~~~~~~~~~~~~~~~~~~~~~~~~~~~~~~~
\begin{abstract}
Stellar population parameters correlate with a range of galaxy properties, but it is unclear which relations are causal and which are the result of another underlying trend. In this series, we quantitatively compare trends between stellar population properties and galaxy structural parameters in order to determine which relations are intrinsically tighter, and are therefore more likely to reflect a causal relation. Specifically, we focus on the galaxy structural parameters of mass $M$, gravitational potential $\Phi\sim M/R_e$, and surface mass density $\Sigma\sim M/R_e^2$. In Barone et~al.\ (2018) we found that for early-type galaxies the age--$\Sigma$ and [Z/H]--$\Phi$ relations show the least intrinsic scatter as well as the least residual trend with galaxy size. In this work we study the ages and metallicities measured from full spectral fitting of 2085 star-forming galaxies from the SDSS Legacy Survey, selected so all galaxies in the sample are probed to one effective radius. As with the trends found in early-type galaxies, we find that in star-forming galaxies age correlates best with stellar surface mass density, and [Z/H] correlates best with gravitational potential. We discuss multiple mechanisms that could lead to these scaling relations. For the [Z/H]--$\Phi$ relation we conclude that gravitational potential is the primary regulator of metallicity, via its relation to the gas escape velocity. The age--$\Sigma$ relation is consistent with compact galaxies forming earlier, as higher gas fractions in the early universe cause old galaxies to form more compactly during their in-situ formation phase, and may be reinforced by compactness-related quenching mechanisms.
\end{abstract}

%~~~~~~~~~~~~~~~~~~~~~~~~~~~~~~~~~~~~~~~~~~~~~~~~~~~~~~~~~~~~~~
\keywords{galaxies: evolution --- galaxies: fundamental parameters ---  galaxies: stellar content --- galaxies: statistics}

%~~~~~~~~~~~~~~~~~~~~~~~~~~~~~~~~~~~~~~~~~~~~~~~~~~~~~~~~~~~~~~
\section{Introduction}\label{intro}

\noindent The stellar population of a galaxy is a cumulative record of the formation and assembly history of its stars. Different stellar population parameters each provide a piece of this complex puzzle. Stellar population age is determined both by when the galaxy first formed stars and how long ago star formation was quenched. Total metallicity [Z/H] tells us about the number of generations of stars the galaxy has formed and whether the current population formed from pristine gas or recycled material. In a complementary fashion, $\alpha$-enhancement [$\alpha$/Fe] provides a measure of star formation duration, by indicating the extent to which the iron, produced in Type Ia supernovae by relatively long-lived stars, is recycled into subsequent stellar populations \citep[e.g.][]{GreggioRenzini1983,Worthey1992,Matteucci1994,PagelTautvaisiene1995,ThomasGreggioBender1998,Thomas2005,McDermid2015}. Using all three of these parameters, we can attempt to reconstruct the broad features of a galaxy's evolutionary history. Understanding what drives changes in these quantities provides insights into the processes shaping galaxy assembly and star formation.

Stellar population parameters have been found to correlate with a wide range of galaxy properties. Many studies have focused on the dependence of stellar population on mass \citep{Gallazzi2005,Gallazzi2006,Gonzalez_Delgado2015,Thomas2010,Lian2018} and velocity dispersion $\sigma$ (early-types: \citealt{Thomas2005,Nelan2005,Robaina2012}; late-types: \citealt{Ganda2007}; early spirals: \citealt{Peletier2007}). Other works have investigated correlations with initial mass function \citep{LaBarbera2013}, morphological type \citep{Ganda2007,Scott2017}, central black hole mass \citep{Martin-Navarro2016}, and structural lopsidedness \citep{Reichard2009}. However it is unclear which (if any) of these correlations imply causation and which are the result of other underlying trends---for example, until recently it was uncertain whether the population--environment relations are causal \citep{Thomas2005,SanchezBlazquez2006,Schawinski2007_environment} or the result of both stellar population and environment correlating with stellar mass $M_*$ \citep{Thomas2010,McDermid2015}. Recent studies by \cite{Liu2016} and \cite{Scott2017} have reconciled this disparity, showing that dependence on mass alone is insufficient to explain observed trends and environment plays a measurable, albeit secondary, role. Furthermore, it is unclear whether the well-studied color-magnitude relation is a consequence of both parameters correlating with $\sigma$ \citep{Bernardi2005} or $M_*$ \citep{Gallazzi2006}. The difficulty is that these trends are often not directly comparable, due to different observational and model uncertainties, and one correlation appearing stronger than another may simply reflect a higher precision in the measurements rather than underlying physics.

By quantitatively comparing scaling relations, several recent studies have demonstrated a clear effect of galaxy size $R_e$ on stellar population for galaxies ranging from highly star forming to quiescent. \cite{Franx2008} found that for massive galaxies out to $z \sim 2$, $M_*$ alone is not a good predictor of star-formation history and that color as a function of stellar mass surface density $\Sigma \propto M_*/R_e^2$ or  gravitational potential $\Phi \propto M_*/R_e$ (referred to as `inferred velocity dispersion') shows less scatter than as a function of $M_*$. This was extended to low redshifts ($z<0.11$) by \cite{Wake2012}, who, by quantifying residual trends when one parameter is held fixed, asserted that $u-r$ color correlates more strongly with $\sigma$ than $\Sigma$, S\'{e}rsic index \citep{Sersic1968}, or $M_*$. Using spectroscopically-derived stellar population parameters for low redshift samples, \cite{Scott2017} and \cite{Li2018} showed that for both early and late-type galaxies much of the scatter in population--mass relations is due to variations with galaxy size, by demonstrating how stellar population varies in the mass--size plane (see also \citealt{McDermid2015} for early-types). Additionally, \cite{vandeSande2018} showed stellar age is tightly coupled with intrinsic ellipticity for both early- and late-type galaxies.

In \citet[hereafter \citetalias{Barone2018}]{Barone2018} we quantitatively compared global stellar population trends in morphologically-identified early-type galaxies by analysing both their intrinsic scatter and residual trends. We focused on the three structural parameters mass $M$, gravitational potential $\Phi \propto M/R_e$, and surface density $\Sigma \propto M/R_e^2$. For each structural parameter we employed two mass estimators: a dynamical mass based on spectroscopic velocity dispersion $\sigma$ and the virial theorem ($M_D \propto \sigma^2 R_e$) and a stellar mass based on photometric luminosity and color ($M_*$). We showed that correlations with $\sigma$ are reproduced using the purely photometric estimator of potential $M_*/R_e$. We found the tightest correlations, and the least residual trend with galaxy size, for the $g-i$ color--$\Phi$, [Z/H]--$\Phi$, and age--$\Sigma$ relations. We found [$\alpha$/Fe] to correlate strongly with both $\Sigma$ and $\Phi$. We concluded that: (1)~the color--$\Phi$ diagram is a more precise tool for determining the developmental stage of a stellar population than the color--$M$ diagram; and (2)~$\Phi$ is the primary regulator for global stellar metallicity, via its relation to the gas escape velocity. The latter is supported by the results of \cite{DEugenio2018}, who showed that gas-phase metallicity in star-forming galaxies is also more tightly correlated with $\Phi$ than either $M$ or $\Sigma$. With regards to the age--$\Sigma$ and [$\alpha$/Fe]--$\Sigma$ correlations, we proposed two possible explanations: either they are the result of compactness-driven quenching mechanisms or they are fossil records of the $\Sigma_{SFR} \propto \Sigma_{gas}$ relation in their disk-dominated progenitors (or some combination of these). To determine which of the various possible physical mechanisms are responsible, we need to know whether these scaling relations are also present in earlier phases of galaxy evolution, while they are still forming stars.

In this paper (Paper~II) we build on the results on stellar populations in early-type galaxies (ETGs) presented in \citetalias{Barone2018} and on gas-phase metallicity in star-forming galaxies (SFGs) by \cite{DEugenio2018}, by studying the ages and metallicities of SFG stellar populations and how they correlate with stellar mass ($M_*$), gravitational potential ($\Phi \propto M_*/R_e$) and surface mass density ($\Sigma \propto M_*/R_e^2$). The overarching approach of this series is to quantitatively compare trends between stellar properties and galaxy dynamics and structure, with the aim of finding the strongest/tightest scaling relations. This paper is arranged as follows. In section~2 we detail the sample selection, and why the dataset used has changed from \citetalias{Barone2018}. Section~3 describes the full spectral fitting method used to measure the stellar population ages and metallicities. In section~4 we present our analysis methods and results for the luminosity-weighted parameters, and in section~5 we present the mass-weighted results. In section~6 we discuss our results and the possible mechanisms responsible, and qualitatively compare to the results presented in \citetalias{Barone2018}. Finally we provide a summary in section~7. Although we perform both luminosity-weighted and mass-weighted fits, we focus predominantly on the luminosity-weighted parameters. Given the galaxies in our sample are star-forming, their spectra are dominated by young stars and so the contribution from low-luminosity old stars is not well constrained, making it difficult to recover the true mass-weighted parameters. Throughout this paper we use the terms `early' and `late' type to refer to a visual morphological classification, whereas `quiescent' and `star-forming' are based on measured star formation rates. While early-type and star-forming are not mutually exclusive categories, we note that the overlap between them is small. Only 7\% of early-types in our sample from \citetalias{Barone2018} would also be classified as star-forming. Therefore for our purposes the categories can be considered disjoint. We assume a $\Lambda$CDM cosmology with $\Omega_M=0.3$, $\Omega_{\Lambda}=0.7$ and $H_0=70$\,km\,s$^{-1}$\,Mpc$^{-1}$, and a \cite{Chabrier2003} initial mass function.

%~~~~~~~~~~~~~~~~~~~~~~~~~~~~~~~~~~~~~~~~~~~~~~~~~~~~~~~~~~~~~~
\section{Sample Selection}

\noindent All data used in this paper is publicly available and based on the SDSS Legacy Survey \citep{York2000,Strauss2002}. An electronic table of the catalog data as well as our derived stellar population parameters is available online, and is described in Table \ref{table}. For our stellar population measurements we use optical spectra from Data Release~7 \citep{Abazajian2009}. We use $r$-band effective radii ($R_e$) from \cite{Simard2011}, as they provide both single and various double S\'{e}rsic fits as well as an $F$-test probability to determine the most appropriate model for each galaxy. To convert from apparent to physical size we use the spectroscopic redshifts given by the SDSS pipeline and assume the standard $\Lambda$CDM cosmology. We use $H_{\alpha}$-derived specific star formation rates \citep[sSFR;][]{Brinchmann2004} from the MPA/JHU catalog, and select star-forming galaxies as having a total sSFR $> 10^{-11.0} \rm{M_{\odot} yr^{-1}}$, and `star forming' locations on the BPT diagram \citep{BaldwinPhillipsTerlevich1981,VeilleuxOsterbrock1987,Kewley2001,Kauffmann2003b,Schawinski2007_AGN} as defined by \cite{Thomas2013}. To ensure reliable stellar population measurements, we select spectra with a median spectral signal-to-noise ratio $\geq$15 per \AA. We use stellar masses ($M_*$) from \cite{Kauffmann2003b} and \cite{Salim2007}, which are derived from spectral energy distribution (SED) fitting. The $M_*$ from \cite{Kauffmann2003b} are based on a \cite{Kroupa2001} initial mass function (IMF), whereas the stellar population models use a \cite{Chabrier2003} IMF. Hence we rescale $M_*$ to a \cite{Chabrier2003} IMF using the conversion from \cite{Madau_Dickinson2014}, $\log M_{\rm Chabrier} = \log M_{\rm Kroupa} - 0.034$.

We compare $M_*$ from \cite{Kauffmann2003b} with $M_*$ derived from our full spectral fits, as well as the $M_*$ derived by \cite{Chang2015} using SDSS spectra and photometry from the Wide-field Infrared Survey Explorer \citep[WISE;][]{Wright2010}, and find good agreement between all three measurements. We prefer to use a partially independent measure of $M_*$ rather than the values derived from our full spectral fits to reduce the correlated errors between $M_*$ and the stellar population parameters. The stellar masses derived by \cite{Chang2015} use the radius measurements by \cite{Simard2011} that we also use in our fits, so to reduce the effect of correlated errors between $M_*$ and $R_e$ artificially tightening the trends, we use $M_*$ from \cite{Kauffmann2003b}. We note, however, that our results are quantitatively unchanged if we instead use the stellar masses from \cite{Chang2015} or from our full spectral fits.

In Paper~I we used a different dataset, namely 625 ETGs from the Sydney-AAO Multi-object Integral-field (SAMI) galaxy survey \citep{Croom2012,Bryant2015,Scott2018}. However, the comparatively extended ongoing star formation in SFGs leads to a higher intrinsic scatter in single-burst parametrizations, so here we require a larger sample than SAMI provides in order to determine the same scaling relations.

%~~~~~~~~~~~~~~~~~~~~~~~~~~~~~~~~~~~~~~~~~~~~~~~~~~~~~~~~~~~~~~
\begin{table*} \begin{center}
\begin{tabular}{ccc} \hline\hline
\textbf{Column Name} & \textbf{Units} & \textbf{Description} \\ \hline
specObjID & ... & SDSS Spectroscopic object ID \\
ObjID & ... & SDSS Photometric object ID \\
Plate & ... & SDSS Plate ID \\
MJD & ... & Modified Julian Date of observation \\
FiberID & ... & SDSS Fiber ID \\
Redshift & ... & SDSS spectrscopic redshift \\
log\_Age\_L & $\log_{10}$ Gyr & Luminosity-weighted age \\
log\_Age\_L\_unc & $\log_{10}$ Gyr & Uncertainty on luminosity-weighted age \\
ZH\_L & ... & Luminosity-weighted total metallicity \\
ZH\_L\_unc & ... & Uncertainty on luminosity-weighted total metallicity \\
log\_Age\_M & $\log_{10}$ Gyr & Mass-weighted age \\
log\_Age\_M\_unc & $\log_{10}$ Gyr & Uncertainty on mass-weighted age \\
ZH\_M & ... & Mass-weighted total metallicity \\
ZH\_M\_unc & ... & Uncertainty on mass-weighted total metallicity \\
log\_Mstar & $\log_{10} M_{\odot}$ & Stellar Mass from Kauffmann et al. 2003 \\
log\_Mstar\_unc & $\log_{10} M_{\odot}$ & Uncertainty on Stellar Mass from \cite{Kauffmann2003b} \\
Re & kpc & Circularised effective radius in r-band from \cite{Simard2011} \\
\rule[-0.5ex]{0pt}{0pt} \\ \hline\hline \end{tabular}  \centering \caption{Description of the table containing our derived stellar population parameters along with the stellar masses from \cite{Kauffmann2003b}, and effective radii from \cite{Simard2011}. This table is available in its entirety in a machine-readable form in the online journal.} \label{tbl}
\end{center} \end{table*}
\label{table}

%~~~~~~~~~~~~~~~~~~~~~~~~~~~~~~~~~~~~~~~~~~~~~~~~~~~~~~~~~~~~~~
\subsection{Aperture Matched Sampling}\label{AMS}

\noindent We employ the technique of Aperture-Matched Sampling (AMS) used by \cite{DEugenio2018}, in which galaxies are selected to have similar physical areas encompassed by the fiber aperture. This technique allows us to mimic the adaptive aperture of integral field surveys while taking advantage of the large and diverse datasets of single-fiber surveys such as the SDSS Legacy Survey. The AMS approach mitigates (at the expense of sample size) the aperture bias inherent to single-fiber surveys that results from probing galaxies over varying areas depending on their apparent size. Combined with radial trends within galaxies, aperture bias can lead to spurious global trends. The aperture-matched subsample is defined by $R_e = R_{\rm fiber}(1 \pm t)$ for some small tolerance $t$. Following \cite{DEugenio2018} we use a tolerance of $13\%$; given the SDSS Legacy Survey fiber radius of 1.5\arcsec, this criterion selects galaxies with 1.3\arcsec$<R_e<$1.7\arcsec. Due to our aperture-matched criterion, our sample has a correlation between galaxy size and redshift. We therefore also require a sample with a narrow range in redshift to remove the effect of our results being due to evolution with redshift  rather than dependence on size. We select galaxies with spectroscopic redshifts $0.043 < z < 0.073$.

\begin{figure}
\includegraphics[width=\columnwidth]{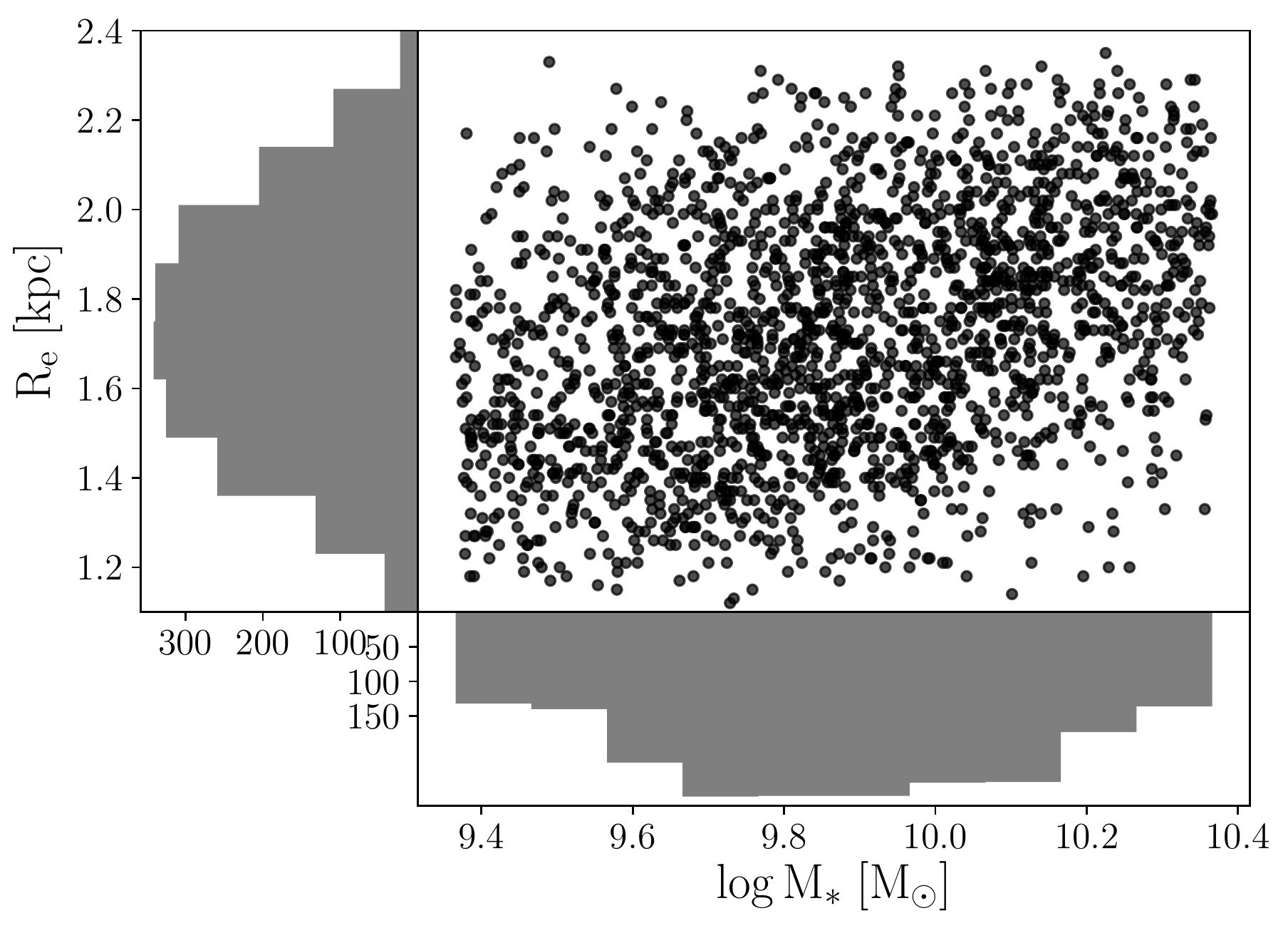}
\caption{Mass--size plane for the sample of SDSS galaxies used here. The sample is approximately evenly distributed in $R_e$ at fixed $M_*$, which reduces bias when  determining the dependence of stellar population parameters on size.}
\label{mass_size_fig}
\end{figure}

%~~~~~~~~~~~~~~~~~~~~~~~~~~~~~~~~~~~~~~~~~~~~~~~~~~~~~~~~~~~~~~
\subsection{Mass-Limited Sample}\label{mass_limited_sample}

\noindent In order to investigate the relative importance of mass and size in predicting stellar population parameters, it helps for the sample to have a similar size distribution at any fixed mass, so there is less in-built mass--size correlation (see Figure~\ref{mass_size_fig}). Consequently we select a mass-limited sample defined by $9.434<\log(M_*/M_\odot)<10.434$. The final sample still has a residual mass-size dependence in that higher mass galaxies have a larger mean size, as removing this completely would severely compromise sample size. While the distribution of sizes at the high and low mass ends of our sample are not identical, the change in the range of sizes is small; the mean size of the galaxies in the lowest and highest mass bins in Figure~\ref{mass_size_fig} (of width 0.1 dex) are 1.58 and 1.88 kpc respectively.

\begin{figure*}
    \includegraphics[width=\textwidth]{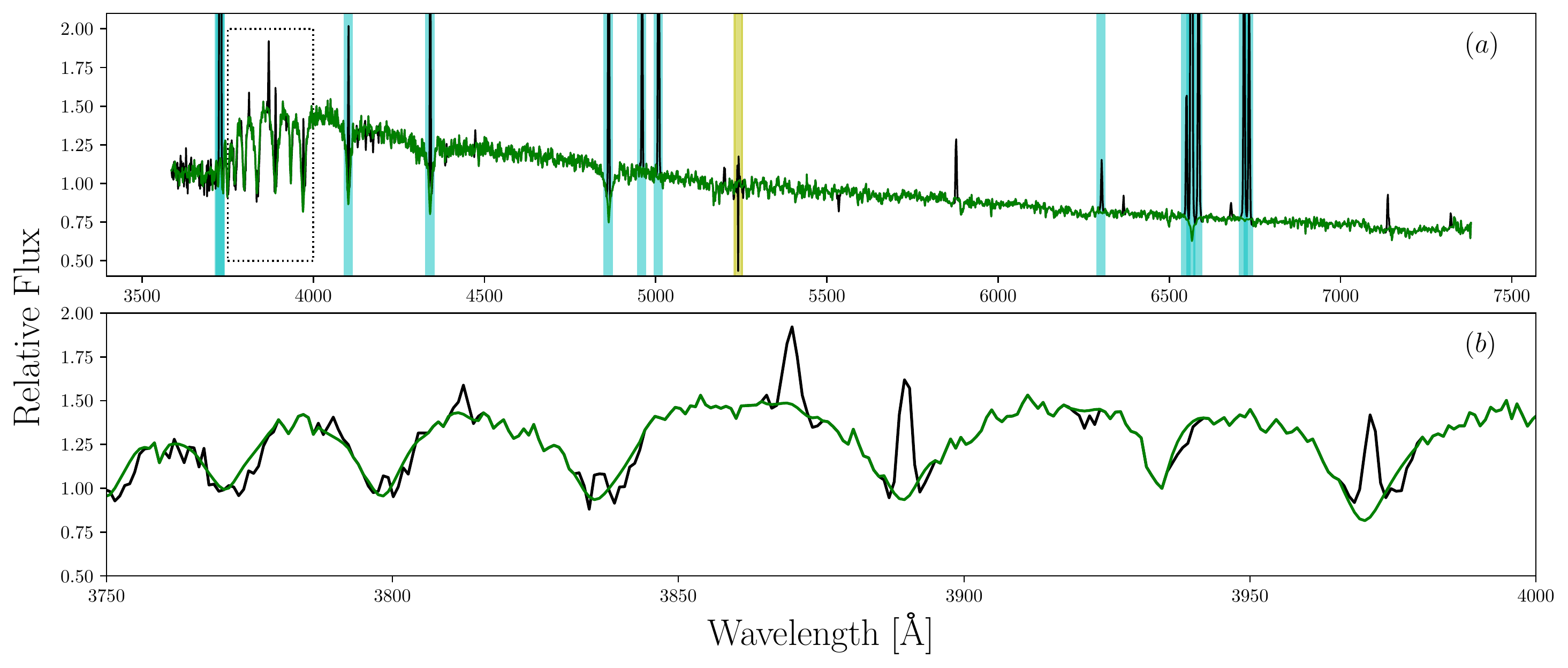}
    \caption{The rest-frame original spectrum for galaxy spec-0541-51959-0600 (black line) and the spectrum used for the stellar population template fitting (green line) that has gas emission lines, sky lines, and discrepant pixels masked. Panel (a) shows the entire wavelength range, panel (b) shows a close-up of the region covering the higher order Balmer lines (indicated by a black dotted box in panel a). The cyan regions are emission lines explicitly masked using the pPXF function \texttt{determine\char`_goodpixels}, and the yellow region is the 5577\AA\ sky line that is also explicitly masked. Any remaining emission lines or discrepant pixels are identified and masked by the CLEAN function in pPXF, which iteratively rejects pixels that deviate more than 3$\sigma$ from the best-fit and refits until no further pixels are clipped \citep{Cappellari2002b}. The higher-order Balmer lines are not explicitly masked, because not all spectra have emission in these regions. However as shown in panel (b), the method used effectively identifies remaining emission lines and masks them, recovering the shape of the underlying absorption feature.}
    \label{spec_fig}
\end{figure*}

%~~~~~~~~~~~~~~~~~~~~~~~~~~~~~~~~~~~~~~~~~~~~~~~~~~~~~~~~~~~~~~
\section{Stellar Population Synthesis}

\noindent We measure the stellar population parameters from full spectral fits using theoretical stellar population models based on the Medium resolution INT Library of Empirical Spectra \citep[MILES;][]{SanchezBlazquez2006_MILES,Vazdekis2010,Vazdekis2015}, using BaSTI isochrones \citep{Pietrinferni2004,Pietrinferni2006} and a \cite{Chabrier2003} initial mass function. This is different from the Lick index method and models used in \citetalias{Barone2018}. The stellar population parameters for \citetalias{Barone2018} were measured by \cite{Scott2017} using the popular Lick system of absorption line indices and the models by \cite{Schiavon2007} and \cite{ThomasMarastonJohansson2011}, as Lick indices afford a benchmark for the analysis of ETG (and globular cluster) populations \citep{Faber1973}. The little-to-no ongoing star formation in ETGs means the spectral absorption lines are free from emission by ionised interstellar gas, allowing for precise measurements. In contrast, SFGs have emission from ionized gas contaminating the absorption features, making it difficult to make accurate measurements. Nevertheless, with high signal-to-noise spectra and careful masking of emission lines, \cite{Ganda2007} and \cite{Peletier2007} were able to find scaling relations with mass and velocity dispersion similar to those found in ETGs. 

However, an alternative is to use sets of theoretical spectra for single-age and single-metallicity populations that allow a full spectral fitting approach using not just a limited number of absorption features but the whole spectrum, including the shape of the continuum. In addition to the MILES models by \cite{Vazdekis2010,Vazdekis2015} used here, other widely used sets of theoretical spectra include \cite{BruzualCharlot2003} and Binary Population and Spectral Synthesis \citep[BPASS;][]{Eldridge2017,StanwayEldridge2018}. While these models do not account for emission from ionized gas, the issue of emission lines obscuring absorption features is less severe with a full spectral fitting method than for Lick indices, because the whole spectrum is used. We therefore use spectral fitting to approach the comparatively less well-studied field of stellar populations in star-forming galaxies.

Despite the different stellar population models used in this work \citep{Vazdekis2010,Vazdekis2015} and in \citetalias{Barone2018}, \citep{Schiavon2007,ThomasMarastonJohansson2011}, there is good agreement between results from these models. \cite{McDermid2015} show that there is a tight relation between stellar population parameters derived using Lick indices and \cite{Schiavon2007} models, and mass-weighted parameters derived from full spectral fitting and \cite{Vazdekis2010} models for ETGs. Their Figure 4 shows that the [Z/H] derived from the two models and methods closely follow the one-to-one relation. The ages follow a tight correlation offset from the one-to-one line, with the luminosity-weighted \cite{Schiavon2007} ages being systematically younger than the  mass-weighted \cite{Vazdekis2010} ages. However this is most likely a result of luminosity-weighted ages being more sensitive to young stars than mass-weighted ages \citep{SerraTrager2007}, rather than a difference in the models used. Additionally, \cite{Scott2017} show that there is good agreement between the \cite{Schiavon2007} and \cite{ThomasMarastonJohansson2011} models, differing most in the low-[Z/H] regime.

Our stellar population analysis consists of two main steps: Step 1 involves masking the spectra of emission and sky lines; Step 2 involves fitting the masked spectrum as a weighted sum of single-age and single-metallicity templates.

%~~~~~~~~~~~~~~~~~~~~~~~~~~~~~~~~~~~~~~~~~~~~~~~~~~~~~~~~~~~~~~
\subsection{Step~1. Emission Line Masking}

\noindent The aim of this pre-processing stage is to mask sky and gas emission lines. We begin by de-redshifting the galaxy and masking known sky and galaxy emission lines. Specifically, we use the function \texttt{determine\char`_goodpixels} from the Python implementation of the publicly available Penalized Pixel-Fitting software (pPXF; \citealt{Cappellari_Emsellem2004,Cappellari2017}) to mask 13 common emission lines (see pPXF documentation for emission line details). Additionally, we also mask the sky line in the region between 5565\AA\ to 5590\AA. These masked regions are highlighted in panel (a) of Figure \ref{spec_fig}. We then perform two fits to the masked spectrum, using pPXF and all 985 empirical stellar templates from the MILES library \citep{SanchezBlazquez2006,Falcon-Barroso2011} broadened to the SDSS instrumental resolution. The purpose of the first fit is to obtain an estimate of the noise and uses the variance given by the SDSS pipeline. Based on the $\chi^{2}_{\rm reduced}$ of the first fit, we then rescale the variances to give $\chi^{2}_{\rm reduced}=1$. The median rescaling value is 0.994 with a standard deviation of 0.079. With this slightly improved noise estimate, the second fit identifies any remaining bad pixels by iteratively rejecting pixels that deviate more than 3$\sigma$ from the best-fit, refitting until no further pixels are rejected (see section~2.1 of \cite{Cappellari2002b} and the CLEAN keyword in the pPXF documentation). Panel (b) of Figure \ref{spec_fig} demonstrates that emission lines not explicitly masked, such as the higher order Balmer lines, are identified and masked by the CLEAN iterative pixel rejection. The pixels identified as bad or containing emission lines are then replaced by the best-fit model from the second fit.

%~~~~~~~~~~~~~~~~~~~~~~~~~~~~~~~~~~~~~~~~~~~~~~~~~~~~~~~~~~~~~~
\subsection{Step~2. Full Spectral Fitting}\label{FSF}

\begin{figure*}
\includegraphics[width=\textwidth]{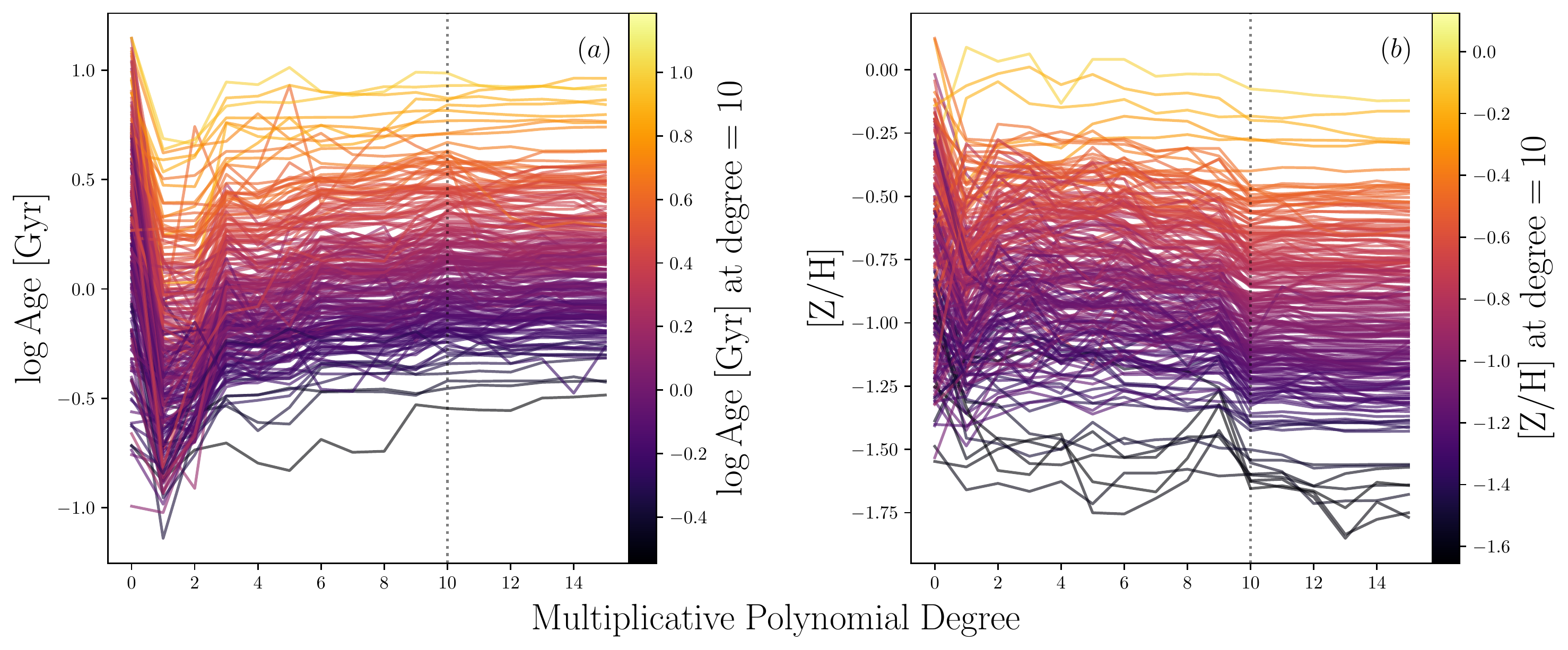}
\caption{The luminosity-weighted age and [Z/H] from fits using varying degrees of the multiplicative polynomial, for a subsample of 209 galaxies. Each line is a single galaxy, and is colored by its age (panel a) and [Z/H] (panel b) from the 10\textsuperscript{th} degree fit. The stellar population parameters vary little above a multiplicative polynomial of degree 10, hence we use a 10\textsuperscript{th} degree polynomial.}
\label{mdegree}
\end{figure*}

\noindent After the pre-processing stage, the stellar population age and metallicity are measured from the masked, emission-line-free spectrum. We fit the masked spectrum as a linear combination of synthetic single-population templates and a degree 10 multiplicative polynomial. The role of the multiplicative polynomial is to correct the shape of the continuum and account for dust extinction, however it significantly increases the computation time \citep{Cappellari2017}. Therefore, ideally the degree of the multiplicative polynomial should be the lowest value such that both residual flux calibration errors and dust extinction are corrected for. Using a randomly selected subsample of 209 galaxies (10\% of the full sample) we tested the dependence of the resulting stellar population parameters on the degree of the multiplicative polynomial used. As shown in Figure \ref{mdegree}, while the absolute values of age and [Z/H] vary greatly for fits with a multiplicative polynomial degree less than 10, the relative difference between galaxies remains similar. The stellar population parameters vary little for degree $\geq 10$, hence we use a degree 10 polynomial. 

The templates used are from \cite{Vazdekis2010,Vazdekis2015} and are constructed from the MILES stellar library and the base [$\alpha$/Fe] BaSTI isochrones \citep{Pietrinferni2004,Pietrinferni2006} and a \cite{Chabrier2003} initial mass function. The base models contain no assumption on the abundance ratios, hence the templates follow the abundance pattern of the Milky Way \citep{Vazdekis2010}. The 636 templates span an approximately regular grid in age and metallicity, spanning $-2.27 \leq [Z/H] \leq 0.40$ ($0.0001 \leq Z \leq 0.040$) and 0.03\,Gyr\,$\leq$\,Age\,$\leq$\,14.0\,Gyr. We perform both luminosity-weighted (i.e. templates are individually normalised; section \ref{luminosity_weighted}) and mass-weighted fits (i.e no renormalization of templates; section \ref{mass_weighted}), however we focus the analysis and discussion predominantly on the luminosity-weighted parameters. Given the galaxies in our sample are star-forming, their spectra are dominated by young stars and so the contribution from low-luminosity old stars is not well constrained, making it harder to recover the true mass-weighted parameters. Each template is assigned a weight and from the combinations of weights a star formation history can be inferred \citep[e.g.][]{McDermid2015}. However, the recovery of the star formation history is an ill-conditioned inverse problem without a unique solution unless further constraints are imposed \citep[e.g.][]{Press2007}. This is because of the not-insignificant degeneracies between stellar spectra with different ages and metallicities. A common solution is to use linear regularization, which constrains the weights of neighbouring templates (in age--metallicity space) to vary smoothly. While linear regularization produces more realistic star-formation histories, typical degrees of regularization (see criterion advocated by \citealt{Press1992} and used by, for example, \citealt{McDermid2015,Norris2015,Boecker2019}) are not expected to significantly change the weighted stellar population parameters, and we confirmed this to be the case for the luminosity-weighted age and [Z/H] using the random subsample of 209 galaxies. However, we did find a small systematic offset between the regularized and non-regularized parameters, in that the regularized values are on average 0.06 dex older and 0.07 dex more metal rich. This offset is introduced by regularizing over templates that are not evenly spaced in age or metallicity. The \cite{Vazdekis2010} templates have larger spacing at older ages, hence smoothing between adjacent templates results in the regularized values being slightly older. This offset is small and less than the median uncertainties on the stellar population parameters ($\rm \sigma_{\log Age} = 0.12$ dex and $\rm \sigma_{[Z/H]} = 0.10$ dex). Overall, we prefer to use the non-regularized fits in estimating the weighted ages and metallicities.

\begin{figure*}
\includegraphics[width=\textwidth]{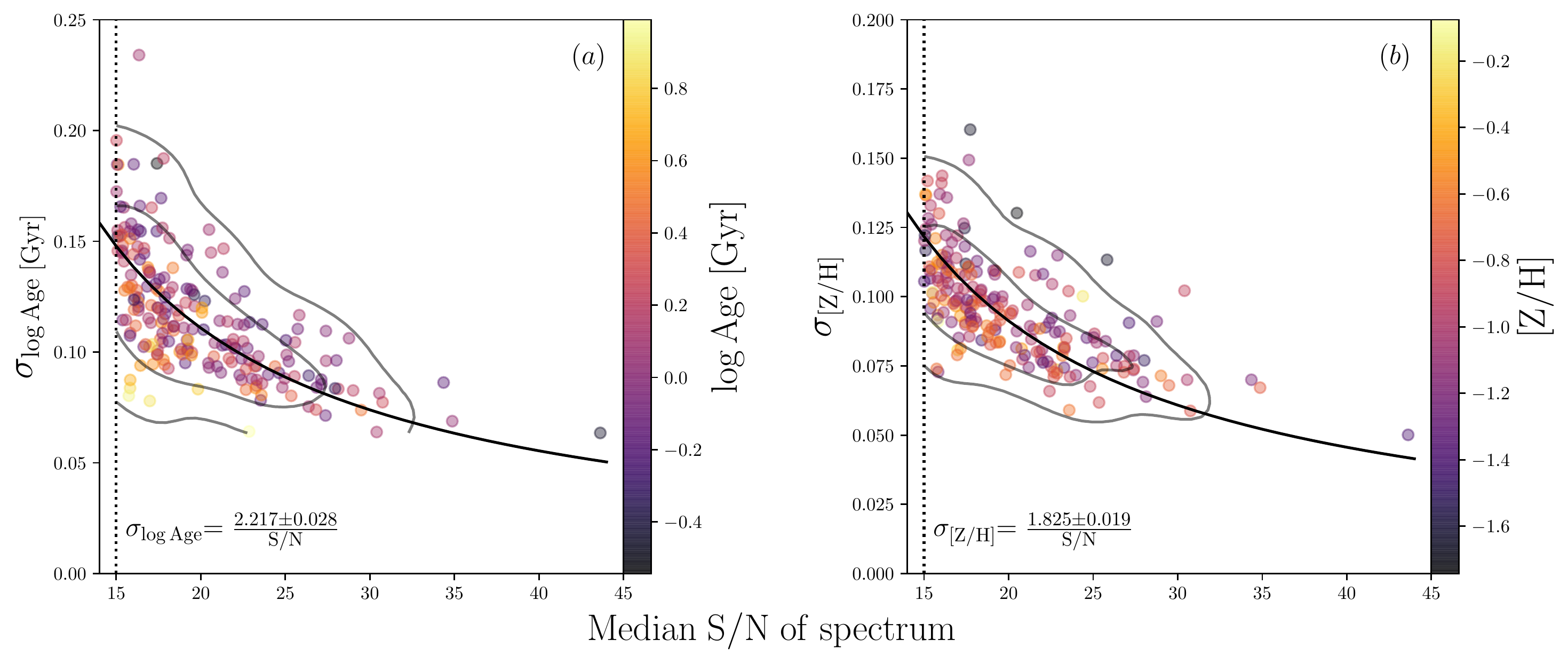}
\caption{Uncertainty on luminosity-weighted age and [Z/H] versus median S/N of the spectrum for the subsample of 209 test galaxies. Each point is colored by its age (panel a) and [Z/H] (panel b). The grey contours enclose 95\% and 68\% of the data. In both panels the black line is the best-fit inverse relation, which is then used to assign an uncertainty on age and [Z/H] to every galaxy in the full sample, based on its spectral S/N.}
\label{uncertainty_SN}
\end{figure*}

%~~~~~~~~~~~~~~~~~~~~~~~~~~~~~~~~~~~~~~~~~~~~~~~~~~~~~~~~~~~~~~
\subsubsection{Estimating Uncertainties}

We derive uncertainties on the luminosity-weighted stellar population parameters as a function of the median S/N per pixel, derived from testing performed on the same 209 galaxies used to test the degree of the multiplicative polynomial. First, we shuffle the residuals from the best-fit within 7 bins approximately 500\AA\ wide, and add this to the best-fit spectrum and refit, repeating 100 times per galaxy. The resulting stellar population parameter distribution is approximately Gaussian and centred around the original fit. Hence we take the standard deviation of the distribution as the uncertainty on the stellar population parameter. Figure \ref{uncertainty_SN} shows the dependence of the measured uncertainty on the median spectral S/N for the 209 test galaxies. Both the uncertainty on age and [Z/H] show an inverse dependence on S/N, which we fit using the Levenberg-Marquardt least-squares optimization algorithm implemented in Python by the \texttt{SciPy} package's \texttt{optimize.curve\char`_fit} routine \citep{scipy}. We then use these relations, $\sigma_{\log \rm Age} = \frac{2.217}{\rm S/N}$, and $\sigma_{\rm [Z/H]} = \frac{1.825}{\rm S/N}$ to assign uncertainties to the age and [Z/H] of each galaxy based on its S/N.

Unlike the luminosity-weighted parameters, the mass-weighted stellar population parameters do not show a strong variation with S/N, and show greater scatter at fixed S/N. Therefore, rather than assigning an uncertainty to each galaxy based on its S/N, we use the median uncertainties from the test subsample, $\sigma_{\log \rm Age} = 0.096$ and $\sigma_{\rm [Z/H]} = 0.18$, and use these values for every galaxy in the sample.

%~~~~~~~~~~~~~~~~~~~~~~~~~~~~~~~~~~~~~~~~~~~~~~~~~~~~~~~~~~~~~~
\section{Luminosity-Weighted Ages and Metallicities}\label{luminosity_weighted}

\begin{figure*}
\includegraphics[width=\textwidth]{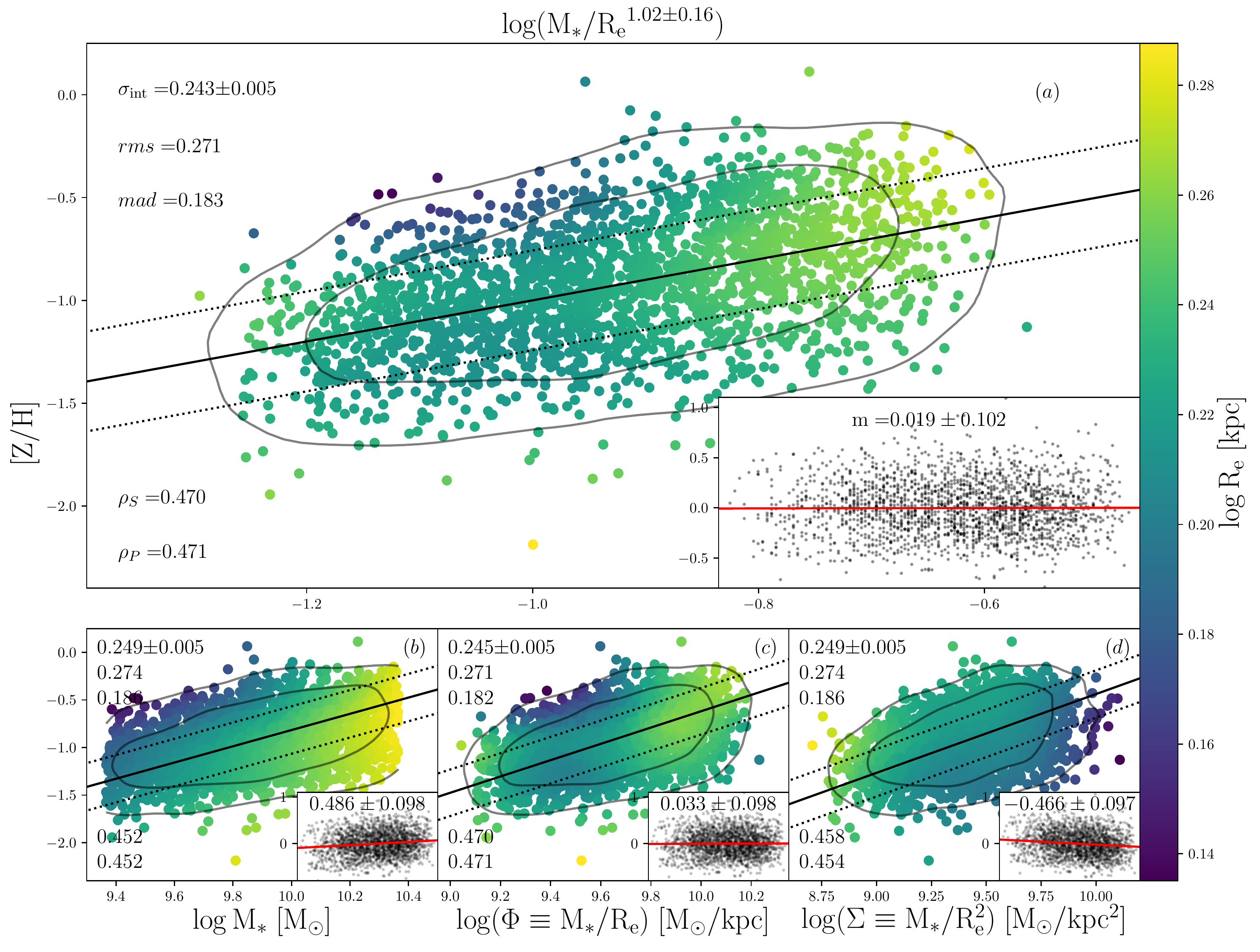}
\caption{Luminosity-weighted [Z/H] versus the best-fit linear combination of $M_*$ and $R_e$ (panel~a) and luminosity-weighted [Z/H] versus $M_*$, $\Phi \equiv M_*/R_e$ and $\Sigma \equiv M_*/R_e^2$ (panels~b--d). In each panel the solid black line is the best-fit linear relation and the dashed lines indicate the intrinsic scatter $\sigma_{\rm int}$ about this fit. The colorscale indicates the LOESS-smoothed value of $\log R_e$~(in kpc). The scatter, both root-mean-square ({\em rms}) and median absolute deviation ({\em mad}), is given at the top left of each panel and the correlation coefficient, both Spearmam $\rho_S$ and Pearson $\rho_P$, is given at the bottom left. The contours enclose 68\% and 95\% of the sample.  The insets show the best-fit residuals versus $\log R_e$; the slope of the residual trend $m$ is displayed at the top of each inset. Panels~(a) and~(c) indicate that of the three structural parameters studied, [Z/H] correlates best with $\Phi$.}
\label{met_fig}
\end{figure*}

%~~~~~~~~~~~~~~~~~~~~~~~~~~~~~~~~~~~~~~~~~~~~~~~~~~~~~~~~~~~~~~
\subsection{Fitting Method}

\noindent We fit both two-parameter lines $z = a_0 + a_1x$ and three-parameter planes  $z=a_0  + a_1x  + a_2y$ to the relationships between stellar population parameters (age and metallicity) and structural properties ($M_*$, $R_e$ and the combinations $\Phi$ and $\Sigma$), allowing for intrinsic scatter in the $z$ direction (i.e.\ in the stellar population parameter). These fits are performed using a Bayesian approach with uniform priors on the slope(s), intercept, and intrinsic scatter. 

The posterior function is first optimised using the Differential Evolution numerical method \citep{StornPrice1997}, followed by Markov Chain Monte Carlo integration \citep{GoodmanWeare2010} of the posterior distribution to estimate the uncertainties on each model parameter using the Python package \texttt{emcee} \citep{Foreman-Mackey2013}.

For both the line and plane fits we quantify the residuals as a function of $R_e$, as displayed in the inset at the bottom right of each panel. In conjunction with the plane fit (where the residual correlation is close to zero by construction), the slopes of the residual correlations illustrate which of $M_*$, $\Phi$ or $\Sigma$ best encapsulates the stellar population parameter's dependence on size.

For each relation we use several metrics to quantify both the significance of the correlation and the tightness of the scatter about the fit. The Spearman and Pearson correlation coefficients ($\rho_S, \rho_P$) characterise the significance, while the root-mean-square ({\em rms}) scatter and median absolute deviation ({\em mad}) about the fit quantify the tightness. The absolute {\em intrinsic} scatter in the relations is difficult to measure, because it is sensitive to the assumed observational uncertainties. However, given the non-zero observational uncertainty on $R_e$, it follows that $M_* R_e^i$ necessarily has a higher total observational uncertainty than $M_*$ alone (for $i \neq 0$). Moreover, if $M_* R_e^j$ shows less scatter than $M_* R_e^i$ for $j > i$,  $M_* R_e^j$ must be intrinsically tighter. Hence, by understanding the relative observational uncertainties, we can compare the measured scatter about the fits and rank the relations based on their relative intrinsic scatter.  The colorscales in the figures show $\log R_e$, smoothed using the locally weighted regression algorithm LOESS \citep{Cleveland_Devlin1988,Cappellari2013b}, to highlight the residual trends with galaxy size.

\begin{figure*}
\includegraphics[width=\textwidth]{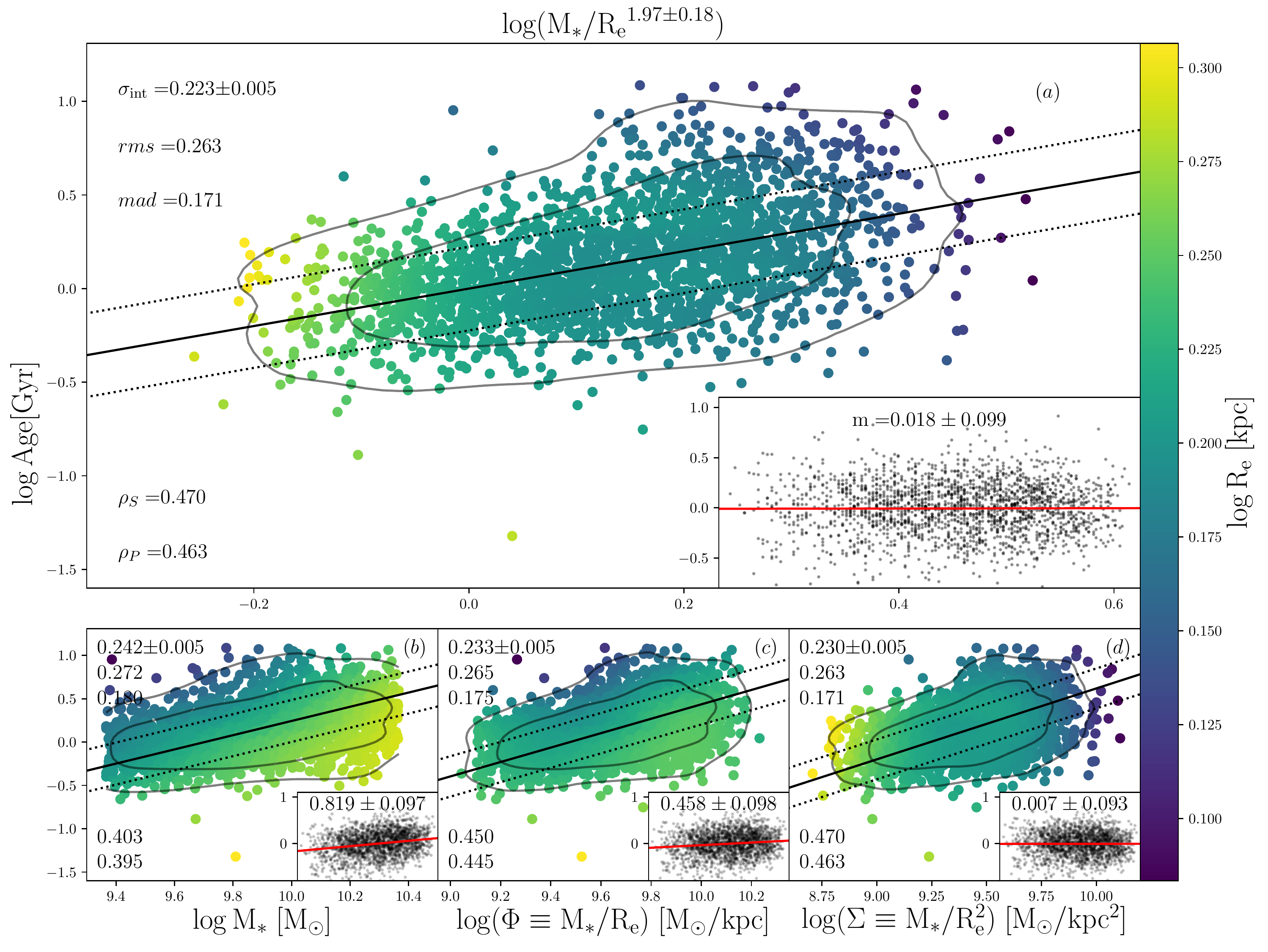}
\caption{Luminosity-weighted age versus the best-fit linear combination of $M_*$ and $R_e$ (panel~a), and luminosity-weighted age versus $M_*$, $\Phi \equiv M_*/R_e$ and $\Sigma \equiv M_*/R_e^2$ (panels~b--d). In each panel the solid black line is the best-fit linear relation and the dashed lines indicate the intrinsic scatter $\sigma_{\rm int}$ about this fit. The colorscale indicates the LOESS-smoothed value of $\log R_e$~(in kpc). The scatter, both root-mean-square ({\em rms}) and median absolute deviation ({\em mad}), is given at the top left of each panel and the correlation coefficient, both Spearmam $\rho_S$ and Pearson $\rho_P$, is given at the bottom left. The contours enclose 68\% and 95\% of the sample. The insets show the best-fit residuals versus $\log R_e$; the slope of the residual trend $m$ is displayed at the top of each inset. Panels~(a) and~(d) indicate that of the three structural parameters studied, age correlates best with $\Sigma$.}
\label{age_fig}
\end{figure*}

%~~~~~~~~~~~~~~~~~~~~~~~~~~~~~~~~~~~~~~~~~~~~~~~~~~~~~~~~~~~~~~
\subsection{Metallicity [Z/H]}

\noindent We show the results of this analysis for [Z/H] in Figure~\ref{met_fig}. Of the three structural parameters, [Z/H]--$\Phi$ in panel~(c) has the tightest correlation. Indeed, the plane fit in panel~(a) shows that the optimum coefficient of $\log R_e$ is $-1.02 \pm 0.16$, consistent within the uncertainties to the $-1$ coefficient corresponding to $\Phi$. Furthermore for the [Z/H]--$\Phi$ relation, the $\sigma_{\rm int}$, {\em rms} and {\em mad} are all consistent within the uncertainties to the plane fit. Moving from left to right in Figure~\ref{met_fig} from $M_*$ (panel~b) through $\Phi$ (panel~c) to $\Sigma$ (panel~d), we see a peak in $\rho_P$ and $\rho_S$, as well as a minimum in $\sigma_{\rm int}$, {\em rms} and {\em mad} at $\Phi$ (panel~c). Invoking the  argument given above, the larger {\em observational} uncertainties in [Z/H]--$\Phi$ compared to [Z/H]--$M_*$, along with slightly less  scatter, implies [Z/H]--$\Phi$ must have less {\em intrinsic} scatter than [Z/H]--$M_*$. 

In addition to the tightness of the fits, the residual trends with $\log R_e$ indicate which of the parameters investigated best encapsulates the dependence of [Z/H] on size. The [Z/H]--$M_*$ diagram in panel~(b) and the [Z/H]--$\Sigma$ diagram in panel~(d) both show significant residual trends with size. As shown by the inset panels, the slopes of the residuals of [Z/H]--$M_*$ and [Z/H]--$\Sigma$ with size are $m = 0.486 \pm 0.098$ and $m = -0.466 \pm 0.097$ respectively. On the other hand, the [Z/H]--$\Phi$ relation shows no residual trend with size ($m= 0.033 \pm 0.098$). This lack of residual trend with size indicates that $\Phi$ best encapsulates the relative dependence of [Z/H] on mass and size.

The results are quantitatively unchanged if we instead use $M_*$ from \cite{Chang2015}. The plane fit using $M_*$ from \cite{Chang2015} is $[Z/H] \propto M_*/R_e^{1.00 \pm 0.13}$, consistent within the uncertainties to our presented results $[Z/H] \propto M_*/R_e^{1.02 \pm 0.16}$ using $M_*$ from \cite{Kauffmann2003b}.

\subsection{Age}

\noindent In Figure~\ref{age_fig}, panels~(b)-(d) show the relations between age and $M_*$, $\Phi$ and $\Sigma$, while panel~(a) shows age fitted by a plane in $M_*$ and $R_e$. For the plane fit, the optimum coefficient of $\log R_e$ is $-1.97 \pm 0.18$, consistent within the uncertainties to the $-2$ coefficient corresponding to $\Sigma$, indicating that despite the high intrinsic scatter and observational uncertainties, age scales most closely with surface mass density $\Sigma$. Indeed the improvement of the plane fit (panel~a) over the age--$\Sigma$ relation (panel~d) is marginal, as indicated by the identical values of $\rho_S$ and $\rho_P$. Moving from left to right in Figure~\ref{age_fig} from $M_*$ (panel~b) through $\Phi$ (panel~c) to $\Sigma$ (panel~d), we see a consistent decrease in the scatter, {\em rms}, {\em mad} and residual slope, along with a corresponding increase in $\rho_P$ and $\rho_S$. Given the higher observational uncertainty of $\Sigma$ compared to $M_*$ or $\Phi$, the tighter correlation with $\Sigma$ implies a fundamentally closer relationship.

Both the age--$M_*$ (panel~b) and age--$\Phi$ (panel~c) relations show significant positive residual trends with size, $m=0.819 \pm 0.097$ and $m=0.458 \pm 0.098$ respectively, whereas the age--$\Sigma$ residuals (panel~d) shows no trend with size ($m=0.007 \pm 0.093$). This lack of residual trend with size indicates that $\Sigma$ best encapsulates the relative dependence of age on mass and size.

If we instead use $M_*$ from \cite{Chang2015} rather than \cite{Kauffmann2003b}, our results remain quantitatively unchanged. The plane fit using $M_*$ from \cite{Chang2015} is age $\propto M_*/R_e^{1.90 \pm 0.16}$, consistent within the uncertainties to age $\propto M_*/R_e^{1.97 \pm 0.18}$ using $M_*$ from \cite{Kauffmann2003b}.

%~~~~~~~~~~~~~~~~~~~~~~~~~~~~~~~~~~~~~~~~~~~~~~~~~~~~~~~~~~~~~~
\section{Mass-Weighted Ages and Metallicities}\label{mass_weighted}

Here we present the mass-weighted stellar population measurements and analyse their dependence on mass and size, to investigate whether the results presented for the luminosity-weighted parameters ($\rm [Z/H]_L$ and $\rm age_L$) in section \ref{luminosity_weighted} hold when using mass-weighted parameters ($\rm [Z/H]_M$ and $\rm age_M$). Unlike $\rm [Z/H]_L$ and $\rm \log age_L$ which show linear dependencies on $\log M_*$, $\log \Phi$, and $\log \Sigma$, both $\rm [Z/H]_M$ and $\rm \log age_M$ show a non-linear dependence on these parameters. We are therefore unable to apply to the mass-weighted parameters the linear fitting method (described in section~\ref{luminosity_weighted}) that we used for the luminosity-weighted parameters. Instead, we analyse the dependence of the mass-weighted parameters on $\log M_*$, $\log \Phi$ and $\log \Sigma$ by showing how $\rm [Z/H]_M$ and $\rm \log age_M$ vary in the mass--size plane.

\begin{figure*}
    \includegraphics[width=\textwidth]{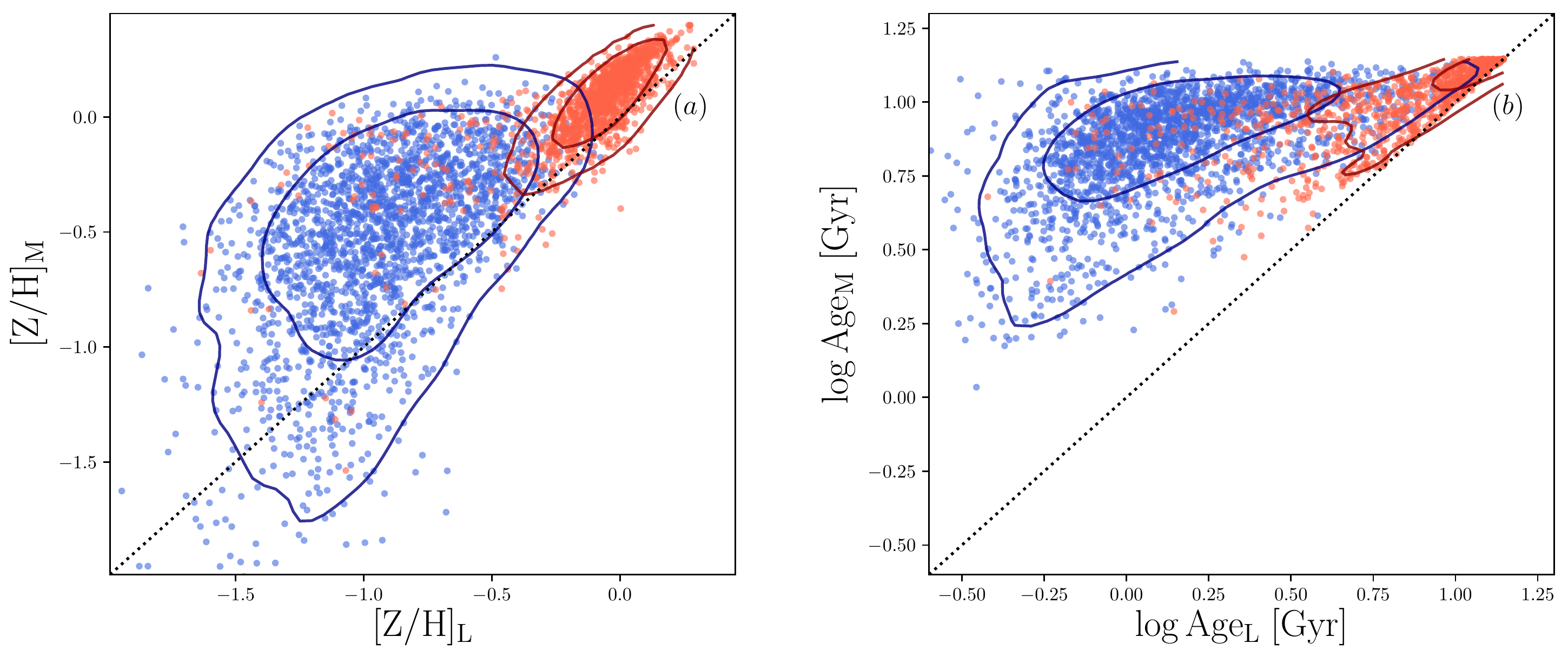}
    \caption{Comparison of mass-weighted and luminosity-weighted [Z/H] (panel~a) and age (panel~b). Blue points are star-forming galaxies,  red points are early-type galaxies from the MOSES catalog. The black dotted line shows the 1-1 relation. The contours enclose enclose 68\% and 95\% of each sample.}
    \label{comparison_mass_light}
\end{figure*}

First, in Figure~\ref{comparison_mass_light} we compare the mass-weighted and luminosity-weighted  parameterizations for both our sample of star-forming galaxies, and an additional sample of early-type galaxies. For the early-types, we use an aperture-matched subsample (following the same criteria described in section~\ref{AMS}) of 1266 galaxies from the MOrphologically Selected Early-types in SDSS \citep[MOSES;][]{Schawinski2007_AGN,Thomas2010} catalog. Similarly, Figure 4 of \cite{McDermid2015} compares mass-weighted [Z/H] and age derived from full spectral fitting to single stellar population (SSP) parameters measured from Lick indices for early-type galaxies from the ATLAS\textsuperscript{3D} survey \citep{Cappellari2011_I}. Given SSP parameters are expected to closely follow luminosity-weighted parameters \citep{SerraTrager2007}, we compare our results with those of \cite{McDermid2015}.

We then show how the the mass-weighted parameters depend on mass and size by how they vary in the mass--size plane (Figure~\ref{met_ms_plane} for [Z/H] and Figure~\ref{age_ms_plane} for age). We include the luminosity-weighted parameters in Figures~\ref{met_ms_plane} and~\ref{age_ms_plane} for reference. To visually highlight the underlying trends we use the LOESS \citep{Cleveland_Devlin1988,Cappellari2013b} algorithm. We compare our luminosity-weighted mass--size planes to similar figures by \cite{Scott2017} and \cite{Li2018}. Specifically, we compare to Figures 9 and 10 of \cite{Scott2017} which show how SSP parameters for SAMI galaxies vary in the mass--size plane, and Figure 4 of \cite{Li2018} who show how luminosity-weighted parameters vary in the mass--size plane for galaxies from the Mapping Nearby Galaxies at APO\citep[MaNGA;][]{Bundy2015} survey.

%~~~~~~~~~~~~~~~~~~~~~~~~~~~~~~~~~~~~~~~~~~~~~~~~~~~~~~~~~~~~~~
\subsection{Metallicity [Z/H]} 

Luminosity-weighted metallicity depends mostly on the old stellar population \citep{SerraTrager2007}, and so we expect good agreement between $\rm [Z/H]_L$ and $\rm [Z/H]_M$. For early-types (red points in Figure~\ref{comparison_mass_light}~a) there is a clear 1-1 relation, although unlike Figure 4. of \cite{McDermid2015} there is a small zero-point offset, with $\rm [Z/H]_M$ being on average $0.16$ dex more metal rich than $\rm [Z/H]_L$. However this 1-1 relation appears limited to above $\rm [Z/H]_L \approx -0.5$, below which there is a significant bend seen strongly in the star-forming galaxies (blue points). In addition to not being 1-1, there is a large variation in $\rm [Z/H]_M$ at fixed $\rm [Z/H]_L$ for the star-forming galaxies.

\begin{figure*}
    \includegraphics[width=\textwidth]{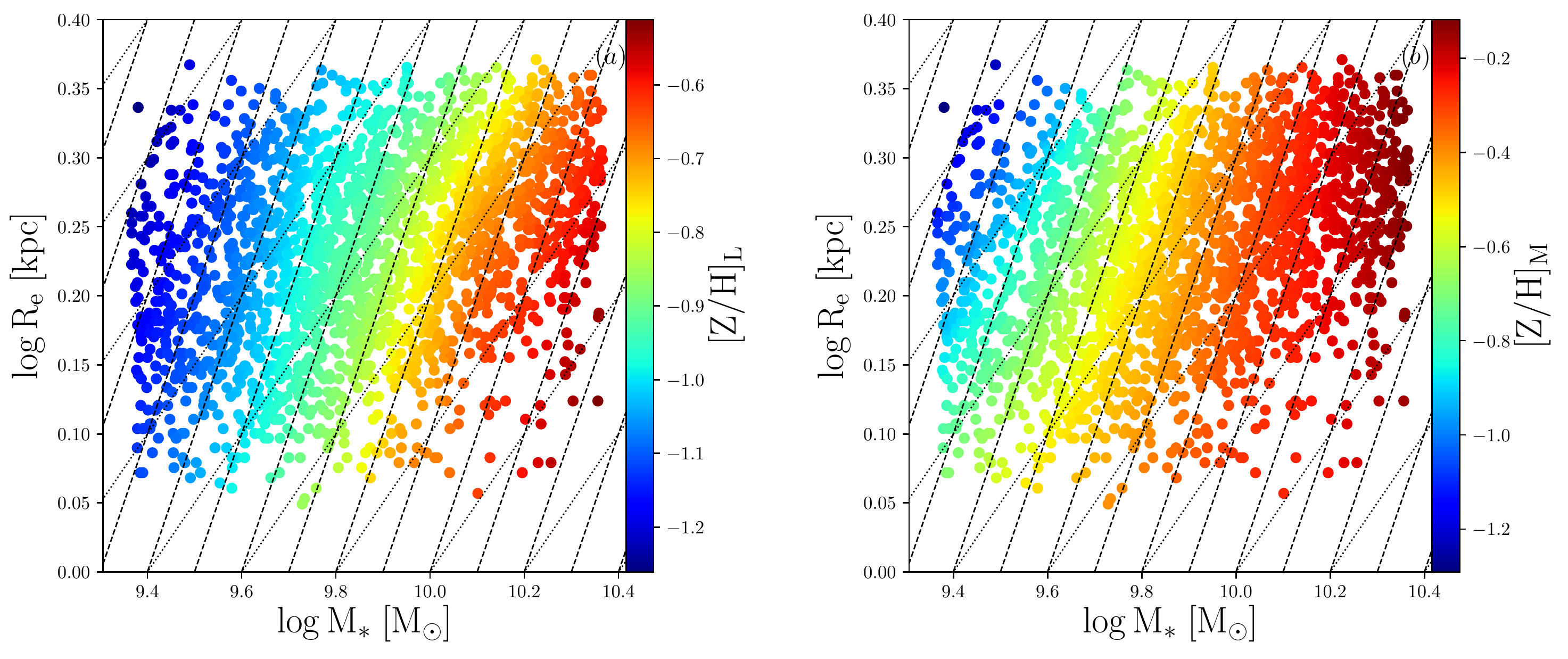}
    \caption{Mass--size plane for our sample of star-forming galaxies, with the colorscale representing LOESS-smoothed luminosity-weighted metallicity ($\rm [Z/H]_L$; panel~a) and mass-weighted metallicity ($\rm [Z/H]_M$; panel~b). The dashed lines are lines of constant $\Phi \propto M_*/R_e$, and the dotted lines are lines of constant $\Sigma \propto M_*/R_e^2$. Both $\rm [Z/H]_M$ and $\rm [Z/H]_L$ follows lines of constant $\Phi$.}
    \label{met_ms_plane}
\end{figure*}

\begin{figure*}
    \includegraphics[width=\textwidth]{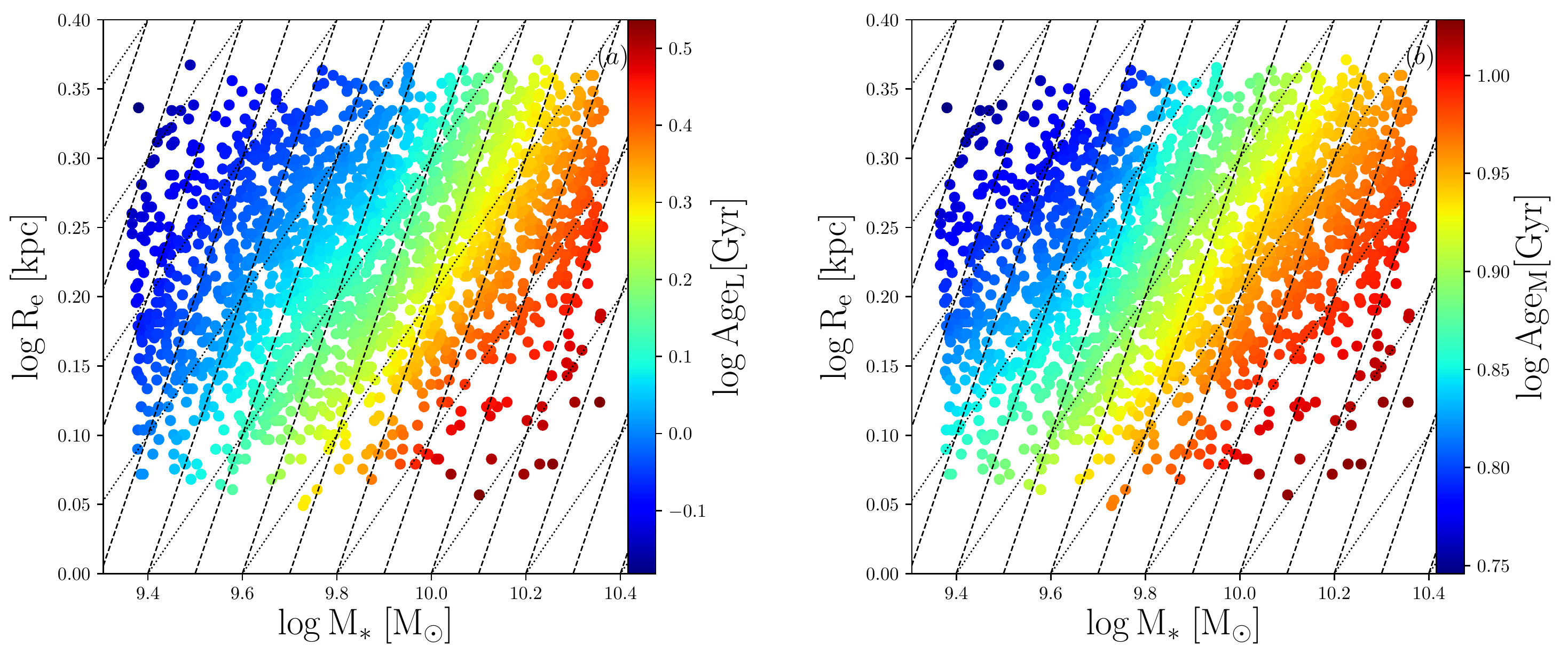}
    \caption{Mass--size plane for our sample of star-forming galaxies, with the colorscale representing LOESS-smoothed luminosity-weighted age ($\rm age_L$; panel~a) and mass-weighted age ($\rm age_M$; panel~b). The dashed lines are lines of constant $\Phi \propto M_*/R_e$ and the dotted lines are lines of constant $\Sigma \propto M_*/R_e^2$. While $\rm age_L$ follows lines of constant $\Sigma$, $\rm age_M$ instead varies somewhere between lines of constant $\Phi$ and $\Sigma$.}
    \label{age_ms_plane}
\end{figure*}

Despite the non-linearity between $\rm [Z/H]_M$ and $\rm [Z/H]_L$, we see in Figure \ref{met_ms_plane} that for star-forming galaxies, $\rm [Z/H]_M$ (panel~b), like $\rm [Z/H]_L$ (panel a), follows lines of constant $\Phi$. \cite{Li2018} also show that $\rm [Z/H]_L$ for spiral galaxies in MaNGA varies along lines of constant $\Phi$, and \cite{Scott2017} show that for all morphological types $\rm [Z/H]_{SSP}$ varies along lines of constant $\Phi$. This strengthens our quantitative results that global stellar metallicity is strongly dependent on the gravitational potential of the galaxy.

%~~~~~~~~~~~~~~~~~~~~~~~~~~~~~~~~~~~~~~~~~~~~~~~~~~~~~~~~~~~~~~
\subsection{Age} 

It is well established that luminosity-weighted ages ($\rm age_L$) strongly trace the younger stars \citep{Trager2000,SerraTrager2007,TragerSomerville2009}, and indeed we see in Figure~\ref{comparison_mass_light} panel (b) that for both the early-types and star-forming galaxies, $\rm age_M$ is consistently older than $\rm age_L$. The early-types (red) resemble the relation shown in Figure~4 of \cite{McDermid2015}, but for the star-forming galaxies (blue) there is a large spread in the $\rm age_M$ at fixed $\rm age_L$. Notably, even at the youngest $\rm age_L$, there are galaxies reaching the upper limit of the templates for $\rm age_M$. For the youngest luminosity-weighted galaxies, it is possible that for these galaxies the spectrum is so dominated by young stars the contribution from low-luminosity old stars is poorly constrained, resulting in over-fitting of the oldest templates. This then leads to the large spread of $\rm age_M$ at fixed $\rm age_L$.

Figure \ref{age_ms_plane} shows that $\rm age_M$ (panel~b) appears to follow lines of constant $\Sigma$, although not as closely as $\rm age_L$. $\rm Age_M$ appears to vary more steeply than $\rm age_L$, at a rate somewhere between lines of constant $\Phi$ and $\Sigma$, although for small, low-mass galaxies (below a stellar mass of $\approx 10^{9.7} M_{\odot}$ and radius $\approx 10^{0.2}$~kpc) $\rm age_M$ appears to closely follow $\Sigma$. \cite{Scott2017} show $\rm age_{SSP}$ also varies approximately along lines of constant $\Sigma$. While \cite{Li2018} do not plot lines of constant $\Sigma$, from their Figure 4 it is clear $\rm age_L$ varies more shallowly than the lines of constant $\Phi$ (lines of constant $\Sigma$ are more shallow than lines of constant $\Phi$).

%~ \\ ~ \\ % force the figures to both be at the top of this page

%~~~~~~~~~~~~~~~~~~~~~~~~~~~~~~~~~~~~~~~~~~~~~~~~~~~~~~~~~~~~~~ 
\section{Discussion}

Our aim was to investigate which parameter (mass $M_*$, gravitational potential $\Phi \sim M_*/R_e$, or surface density $\Sigma \sim M_*/R_e^2$) best predicts the stellar population properties (age and metallicity) of star-forming galaxies. Looking both at the luminosity-weighted (section \ref{luminosity_weighted}) and mass-weighted (section \ref{mass_weighted}) parameters and taking into account both the tightness of the relations and any residual trends with galaxy size, we find age correlates best with surface density while metallicity [Z/H] correlates best with gravitational potential. These results are in striking agreement with \citetalias{Barone2018}, where, using different methods to determine stellar population parameters, we found early-type galaxies also show age correlating best with $\Sigma$ and [Z/H] correlating best with $\Phi$. We note that `quiescent/star-forming' refers to a classification based on specific star formation rate whereas `early-type/late-type' refers to a morphological classification, so `early-type' and `star-forming' are not mutually exclusive categories (see section \ref{intro} for further discussion of the overlap). However given the pronounced differences in internal structure, kinematics, and stellar population properties between the two categories, it is significant that they exhibit the same scaling relations. Crucially, this indicates that the dominant mechanism(s) driving stellar population evolution must originate while galaxies are still star-forming, and must be (at least) preserved through mergers and quenching processes. Here we discuss various mechanisms that could lead to these scaling relations.

%~~~~~~~~~~~~~~~~~~~~~~~~~~~~~~~~~~~~~~~~~~~~~~~~~~~~~~~~~~~~~~
\subsection{Origin of the Metallicity--Potential Relation}

\noindent We have demonstrated that global {\em stellar} metallicity exhibits a tight correlation with the gravitational potential for both early-type galaxies (\citetalias{Barone2018}) and star-forming galaxies (this paper). \cite{DEugenio2018} found the same result for global {\em gas-phase} metallicity in star-forming galaxies. Furthermore, recent works hint at the existence of a similar global relation at higher redshift. \cite{Diaz-Garcia2019} showed that at $z \sim 1$ more compact quiescent galaxies are both older and more metal-rich than their diffuse counterparts at fixed mass.

In general, total metal content is a reflection of the number of generations of stars the galaxy has formed. However, we can rule out the [Z/H]--$\Phi$ relation being driven simply by the number of stellar generations, due to the existence of a strong correlation between the star formation duration (via either [$\alpha$/Fe] or the e-folding timescale) and gravitational potential in both early-types ($M/R_e$: \citealt{Barone2018}; $\sigma$: \citealt{Thomas2005,Nelan2005,Thomas2010,Robaina2012,McDermid2015,Scott2017}) and late-types \citep{Ganda2007}, since galaxies with a shallower potential (lower $\sigma$) have {\em longer}, rather than shorter, star formation durations.

The existence of the [Z/H]--$\Phi$ relation in both the gas and stars, in both young star-forming galaxies and old early-types, indicates the relation originates with in-situ star-formation, and is maintained throughout ex-situ assembly. Although the radius to which we probe ($\sim$1$R_e$) is dominated by in-situ stars \citep{Pillepich2014,Cook2016,Greene2019}, we explore mechanisms related to both in- and ex-situ formation to explain the presence of the metallicity--potential relation. Regarding in-situ formation, either low-$\Phi$ galaxies {\em lose a higher fraction} of their metals or low-$\Phi$ galaxies {\em produce less} metals. In the following discussion we explore two possibilities, namely: (1)~low-$\Phi$ galaxies are more likely to lose more metals, due to the relation between gravitational potential and gas escape velocity; or (2)~low-$\Phi$ galaxies produce less metals due to variations in the initial-mass function. We then discuss how the relation could be preserved in galaxy mergers.

%~~~~~~~~~~~~~~~~~~~~~~~~~~~~~~~~~~~~~~~~~~~~~~~~~~~~~~~~~~~~~~

\subsubsection{Metallicity is determined by gas escape velocity?}

In \citetalias{Barone2018} we proposed that the metallicity--potential relation is driven by low-$\Phi$ galaxies being more likely to lose their metals due to the relation between gravitational potential and gas escape velocity. The depth of the gravitational potential sets the escape velocity for ejection from the galaxy for metal-rich gas expelled by supernovae. This dependence of the gas escape-velocity on the gravitational potential also explains the existence of metallicity gradients within galaxies: the gravitational potential decreases outwards in galaxies, allowing stellar feedback to more easily eject metals in the outskirts than in the centre \citep{Cook2016} and leading to decreasing radial stellar metallicity gradients, as observed in both late \citep{Sanchez-Blazquez2014} and early-type galaxies \citep[e.g][]{Ferreras2019,Goddard2017,Martin-Navarro2018}. This interpretation is supported by the results of \cite{Scott2009}, who found a strong correlation in early-types between {\em local} [Z/H] and {\em local} escape velocity derived from dynamical models. \cite{MollerChristensen2019} also show that halo gas-phase metallicities are well explained by a dependence of the local gas-phase metallicity on the local gravitational potential. Supporting this explanation, simulations show that steep stellar population gradients are the result of in-situ star formation \citep{Pipino2010}, and mergers then tend to diminish these gradients \citep{Kobayashi2004,DiMatteo2009}, particularly at large radii where the stars have predominantly ex-situ origins \citep{Hirschmann2015}. 

A test of this hypothesis is how the metallicity of the circumgalactic medium (CGM) correlates with galaxy structure; logically this mechanism should lead to a relative enrichment of the CGM around low-$\Phi$ galaxies at fixed $M_*$. Due to the low density nature of the CGM, obtaining precise metallicity measurements is time-consuming, and recent studies have sample sizes of less than 50 galaxies \citep[e.g.][]{Prochaska2017,Pointon2019}. In addition, because the CGM is composed not only of stellar ejecta but also pristine gas from the halo and low-metallicity gas from satellites \citep[e.g.][]{Shen2013}, any trend with galaxy gravitational potential would be difficult to interpret. An alternative way forward might be to investigate the dependence of CGM metallicity on galaxy structure in large-scale cosmological simulations of galaxy formation. 

%~~~~~~~~~~~~~~~~~~~~~~~~~~~~~~~~~~~~~~~~~~~~~~~~~~~~~~~~~~~~~~

\subsubsection{Metallicity is determined by initial mass function?}

Another explanation for low-$\Phi$ galaxies producing fewer metals could be variations in the types of stars formed, i.e.\ the initial mass function (IMF). Different stellar types produce different chemical yields and, combined with their varying lifespans, affect both total metallicity [Z/H] and $\alpha$-enhancement, with higher-mass stars leading to higher metallicities and $\alpha$-enhancements \citep[see e.g.][]{Matteucci2012}. Indeed, \cite{Vincenzo2016} showed that the more top-heavy IMFs \citep{Kroupa2001,Chabrier2003}, with their greater proportion of high-mass stars, lead to twice the oxygen yields of the standard \cite{Salpeter1955} IMF. Furthermore there is mounting evidence for a varying IMF both between \citep[e.g.][]{vanDokkumConroy2010,Conroy_vanDokkum2012,Cappellari2012,Spiniello2014,Li2017} and within galaxies \citep{Martin-Navarro2015b,vanDokkum2017,Vaughan2018,Parikh2018}, although exactly what drives these variations remains unclear. 

On the other hand, metallicity has been suggested to {\em anti}-correlate with the relative number of high-mass stars, both globally \citep{Marks2012} with [Fe/H] and locally \citep{Martin-Navarro2015c} with total metallicity [Z/H]. However in contrast, recent works have found that while both metallicity and IMF vary radially, spatially resolved maps show that IMF variations do not follow total metallicity [Z/H] variations \citep{Martin-Navarro2019}. Given these results, while the IMF clearly plays an important role in overall metal production, we find IMF variations do not explain the global metallicity--potential relation.

\subsubsection{Ex-situ preservation}

\noindent In addition to the previously discussed {\em generative} in-situ mechanisms, in order for the metallicity--potential relation to persist in ETGs it must be {\em preserved} during galaxy mergers. While simulations show that mergers tend to diminish metallicity gradients \citep{Kobayashi2004}, it is possible that the global relation is preserved due to the compactness of a satellite influencing where it accretes onto the host. Using N-body simulations, both \cite{Boylan-Kolchin_Ma2007} and \cite{Amorisco2017} show that a compact, high-density satellite is more likely to accrete into the centre of the host, whereas a diffuse, low-density satellite is more easily disrupted by dynamical friction and therefore accretes onto the host's outskirts. This differential process acts to reinforce the already established in-situ metallicity--potential relation: compact, high-$\Phi$ satellites will have relatively high metallicity and deposit their high-metallicity material into the centre of the host, increasing the host's gravitational potential. Conversely, diffuse, low-$\Phi$ satellites will deposit low-metallicity material at large radii, decreasing the host's gravitational potential at fixed mass. Additionally, \cite{Scott2013} find that, despite their different merger histories, both fast and slow rotating early-type galaxies lie on the same scaling relation between the Mg$b$ spectral index and local escape velocity $V_{\rm esc}$. They show that simple model parameterisations indicate dry major mergers should move galaxies off, not along, the relation, and so the intrinsic scatter in the relation therefore provide an upper estimate on the frequency of dry major mergers. Combining predictions from N-body binary mergers and the observed scatter about the Mg$b$--$V_{\rm esc}$ relation, they estimate a typical present-day early-type galaxy to have typically only undergone about 1.5 dry major mergers.

Future studies comparing the slope of the metallicity--potential relation over all galaxy types, at low and high redshift, could further reveal the relative importance of these in- and ex-situ mechanisms, and the precise extent to which mergers diminish or preserve the relation. 

%~~~~~~~~~~~~~~~~~~~~~~~~~~~~~~~~~~~~~~~~~~~~~~~~~~~~~~~~~~~~~~

\subsection{Origin of the Age--$\Sigma$ Relation}

\noindent We find stellar age correlates best with surface mass density $\Sigma$ for both star-forming and early-type galaxies (\citetalias{Barone2018}). While the true average stellar population age depends on when the galaxy first formed, the rate of star formation, and when the galaxy quenched, in practice single-burst model ages strongly depend on the age of the youngest stars \citep{Trager2000,SerraTrager2007,TragerSomerville2009}. In \citetalias{Barone2018} we proposed two possible explanations for the age--$\Sigma$ relation in ETGs: (1)~as a fossil record of the $\Sigma_{\rm SFR} \propto \Sigma_{\rm gas}$ relation while forming stars, or (2)~as a result of compactness-driven quenching mechanisms.

For ETGs these two scenarios are completely degenerate, but in this work, because we use star-forming galaxies, we are able to break this degeneracy. In fact, given the result of this paper that the age--$\Sigma$ relation also exists in SFGs, it would be an odd coincidence if the same relation was due to completely different physical processes. Assuming therefore the mechanism(s) leading to this relation is (are) the same for ETGs and SFGs, the relation must originate {\em before} quenching. Nonetheless, certain quenching mechanisms may further emphasize the relation. Here we discuss mechanisms related to each of these phases that could lead to or reinforce the age--$\Sigma$ relation. Firstly we explore whether galaxies that formed earlier have high-$\Sigma$ due to higher gas densities in the early universe, building upon the hypothesis from \citetalias{Barone2018} that the relation is a fossil record of the $\Sigma_{\rm SFR} \propto \Sigma_{\rm gas}$ relation. We then discuss the possibility that compact galaxies quench earlier.

%~~~~~~~~~~~~~~~~~~~~~~~~~~~~~~~~~~~~~~~~~~~~~~~~~~~~~~~~~~~~~~
\subsubsection{Compact galaxies formed earlier?}\label{formed_earlier}

\noindent The age--$\Sigma$ relation could be a result of more compact galaxies having formed earlier, because higher gas fractions in the early universe mean galaxies formed more compactly during their in-situ formation phase \citep{Wellons2015}. While this mechanism would apply to both SFGs and ETGs, we first consider the body of evidence related to ETGs, then consider how this also affects SFGs.

The current paradigm from both observations and simulations is that present-day ETGs underwent two main phases of evolution: an early period of intense in-situ star formation at $z\sim$2, producing the very compact galaxies observed at high redshift \citep[e.g.][]{vanDokkum2008,vanderWel2008}, followed by passive ex-situ build-up via frequent minor and occasional major mergers \citep[e.g.][]{Oser2010,Barro2013,Rodriguez-Gomez2016,Wellons2016}. During the in-situ phase the high gas density leads to a high star formation rate density, a causation parameterized by the Kennicut-Schmidt relation (\citealt{Schmidt1959,Kennicutt1998}; see \citealt{KennicuttEvans2012} for a review). As previously discussed in \citetalias{Barone2018}, the Kennicut-Schmidt relation, $\Sigma_{\rm SFR} \propto \Sigma_{\rm gas}$, in SFGs naturally leads to an age--$\Sigma_*$ relation. A high gas density causes a high star formation rate (SFR) density and, assuming a non-replenishing gas supply, quickly exhausts the available gas, leading to a short star-formation duration and an old stellar population. Over time, the original high gas density is converted into a high stellar mass density. In addition, \cite{Tacconi2013} show that the Kennicutt-Schmidt relation is near-linear from redshifts $z \sim 1-3$, indicating this affects both old and young galaxies. Indeed, \cite{Franx2008} showed that specific star formation rate tightly anti-correlates with surface mass density (tighter than mass alone), concluding that star formation history is strongly dependent on surface mass density. \citetalias{Barone2018} also showed that [$\alpha$/Fe], a proxy for star formation duration, correlates tightly with $\Sigma$ (and $\Phi$). Although still star forming, this fossil record of $\Sigma_{\rm SFR} \propto \Sigma_{\rm gas}$ is already detectable in our sample of SFGs as the age--$\Sigma$ relation. Given the mass range of our sample of SFGs, $10^{9.4} < M_* / M_\odot < 10^{10.4}$, enough of the galaxies' star-forming period has passed for the relation with stellar age to be detectable. While the luminosity-weighted ages of SFGs are young, as discussed in section 5.2, luminosity-weighted ages predominantly trace the youngest stars, and the stellar population overall is likely much older as indicated by the mass-weighted ages, which are significantly older. 

Additionally, at low redshift, SFGs are larger than quiescent galaxies at fixed mass \citep{Shen2003,Trujillo2007,Cimatti2008,Kriek2009,Williams2010,Wuyts2011,vanderWel2014,Whitaker2017}, indicating that currently star-forming galaxies are different from the progenitors of present-day compact quiescent galaxies, and will evolve into extended quiescent galaxies \citep{Barro2013}. This explains both why old SFG are more compact than young SFG, and also why early-types are more compact than SFGs. In this scenario, the age--$\Sigma$ relation is a reflection of the gas density of the universe when the galaxy formed. We note, however, that any mechanism that causes a high gas density would also produce an age--$\Sigma$ relation.

%~~~~~~~~~~~~~~~~~~~~~~~~~~~~~~~~~~~~~~~~~~~~~~~~~~~~~~~~~~~~~~
\subsubsection{Compact galaxies quench earlier?}

\noindent In \citetalias{Barone2018} we proposed that the age--$\Sigma$ relation in ETGs might be a result of compactness-driven quenching mechanisms. However, given the result of this work that age correlates tightly with $\Sigma$ also for star-forming galaxies, we assume the mechanism(s) leading to the age--$\Sigma$ relation is (are) the same for both quiescent and star-forming galaxies. Therefore, we infer the relation arises {\em before} quenching, thus disfavouring models where the relation is purely due to quenching. Nonetheless, quenching processes may act to reinforce an already-existing relation\footnote{In principle, quenching could still be responsible for the observed trend if most SFGs had undergone a quenching phase, followed by rejuvenation. In practice, however, rejuvenation is not common and most SFGs have extended star-formation histories \citep[e.g.][]{Thomas2010,Chauke2019}}. Further work quantitatively comparing the age--$\Sigma$ relation in low redshift samples of quiescent and SFGs may help resolve whether the relations originate from the same mechanism(s).

Star formation history and quiescence, as quantified in a variety of ways, correlate strongly with compactness and the presence of a central bulge, both at low \citep{Kauffmann2003a,Kauffmann2004,Bell2008,Franx2008,vanDokkum2011,Bluck2014,Omand2014,Woo2017} and high redshifts \citep[e.g.][]{Bell2012,Wuyts2011,Cheung2012,Szomoru2012,Lang2014}. \cite{Woo2015} proposed two main quenching pathways which may act concurrently: rapid central compactness-related processes, and prolonged halo (environmental) quenching. Compactness-related quenching mechanisms include processes which both build the central bulge and (either directly or indirectly) contribute to quenching, such as mergers and gaseous inflows from the disk. Specifically, gaseous inflows from the disk to the bulge, triggered by disk instability or an event such as a merger, are exhausted in a starburst, leading to increased bulge compactness. Additionally, these inflows can trigger active galactic nuclei which, if aligned with the gas disk, can cause molecular outflows, depleting surrounding gas on timescales of a few Myr and preventing further star formation \citep{Garcia-Burillo2014,Sakamoto2014}. More recently, \cite{WooEllison2019} showed that, in addition to these compactness-related mechanisms, processes unrelated to central density such as secular inside-out disk growth \citep{LillyCarollo2016} combined with slow environmental quenching also naturally lead to a relation between the compactness of the galaxy (which they define by the surface mass density within the central 1kpc, $\Sigma_{\rm 1kpc}$) and quiescence (defined by low sSFR). This compactness--quiescence relation would then naturally lead to a relation between surface mass density and stellar age in passive galaxies.

%~~~~~~~~~~~~~~~~~~~~~~~~~~~~~~~~~~~~~~~~~~~~~~~~~~~~~~~~~~~~~~
\section{Summary}

In this work we have used 2- and 3- dimensional fits to study how the age and metallicity [Z/H] of the global stellar population in star-forming galaxies correlate with the galaxy structural parameters stellar mass ($M_*$), gravitational potential ($\Phi\sim M_*/R_e$), and surface mass density ($\Sigma \sim M_*/R_e^2$). This new study builds on our results for early-type galaxies (\citetalias{Barone2018}). For both early-type and star-forming galaxies, we find the tightest correlations and least residual trend with galaxy size for the age--$\Sigma$ and [Z/H]--$\Phi$ relations. Finding these trends in both these studies, despite the different samples, methods, and models used to derive not only the stellar population parameters but also the stellar masses and effective radii, suggests our results are robust. We discuss multiple mechanisms that might produce these relations. We suggest that the [Z/H]--$\Phi$ relation is driven by low-$\Phi$ galaxies losing more of their metals because the escape velocity required by metal-rich gas to be expelled by supernova feedback is directly proportional to the depth of the gravitational potential. This relation is preserved during mergers, as elucidated by simulations. We rule out the possibility of the [Z/H]--$\Phi$ relation being due to IMF variations. In \citetalias{Barone2018} we discussed compactness-related quenching mechanisms which could lead to the age--$\Sigma$ relation, however given in this work we show that the relation exists also in star-forming galaxies, it must arise {\em before} quenching. We therefore explore the possibility that the age--$\Sigma$ relation is a result of compact galaxies forming earlier. Additionally, certain compactness-related quenching mechanisms may act to reinforce the already-existing relation. Future studies using cosmological simulations may help resolve the relative importance of each of these mechanisms.

%~~~~~~~~~~~~~~~~~~~~~~~~~~~~~~~~~~~~~~~~~~~~~~~~~~~~~~~~~~~~~~
\section{Acknowledgements}

T.M.B.\ is supported by an Australian Government Research Training Program Scholarship. N.S. acknowledges support of an Australian Research Council Discovery Early Career Research Award (project DE190100375) funded by the Australian Government. This research was supported by the Australian Research Council Centre of Excellence for All Sky Astrophysics in 3 Dimensions (ASTRO 3D), through project CE170100013. FDE acknowledges funding through the H2020 ERC Consolidator Grant 683184. We thank the referee for constructive comments.

We make extensive use of the Python programming language and IPython extension \citep{IPython} and packages NumPy \citep{NumPy}, SciPy \citep{scipy}, Astropy \citep{Astropy} and matplotlib \citep{Hunter2007}. In preliminary analyses, we also used TOPCAT \citep{Taylor2005}.

Funding for the SDSS and SDSS-II has been provided by the Alfred P. Sloan Foundation, the Participating Institutions, the National Science Foundation, the U.S. Department of Energy, the National Aeronautics and Space Administration, the Japanese Monbukagakusho, the Max Planck Society, and the Higher Education Funding Council for England. The SDSS Web Site is http://www.sdss.org/.

The SDSS is managed by the Astrophysical Research Consortium for the Participating Institutions. The Participating Institutions are the American Museum of Natural History, Astrophysical Institute Potsdam, University of Basel, University of Cambridge, Case Western Reserve University, University of Chicago, Drexel University, Fermilab, the Institute for Advanced Study, the Japan Participation Group, Johns Hopkins University, the Joint Institute for Nuclear Astrophysics, the Kavli Institute for Particle Astrophysics and Cosmology, the Korean Scientist Group, the Chinese Academy of Sciences (LAMOST), Los Alamos National Laboratory, the Max-Planck-Institute for Astronomy (MPIA), the Max-Planck-Institute for Astrophysics (MPA), New Mexico State University, Ohio State University, University of Pittsburgh, University of Portsmouth, Princeton University, the United States Naval Observatory, and the University of Washington.

%~~~~~~~~~~~~~~~~~~~~~~~~~~~~~~~~~~~~~~~~~~~~~~~~~~~~~~~~~~~~~~
\bibliography{bibfile.bib}

\end{document}